\documentclass{PoS}
\usepackage{mathtools}

\DeclarePairedDelimiter\ket{\lvert}{\rangle}
\DeclarePairedDelimiterX\braket[2]{\langle}{\rangle}{#1 \delimsize\vert #2}

\def\Journal#1#2#3#4{{#1} {\bf #2}, #3 (#4)}

\def\iiser{Department of Physical Sciences\\
Indian Insitute of Science Education and Research, Mohali, Punjab INDIA}
\def\support{On behalf of the Belle Collaboration, supported by the Department of Science and Technology, India}
\def\PLB{{ Phys. Lett.}  B}
\def\PRL{ Phys. Rev. Lett.}
\def\PRD{{ Phys. Rev.} D}

\title{Latest results on mixing and $CP$ violation in the charm decays at the $B$-factories}

\ShortTitle{Latest charm mixing and $CP$ results from $B$-factories}

\author{\speaker{Vishal Bhardwaj }\thanks{\support.}\\
        \iiser\\
        E-mail: \email{vishstar@gmail.com}}

\abstract{In this report, latest results on mixing and $CP$ violation in the 
  charm decays at the $B$-factories are presented. }

\FullConference{9th International Workshop on the CKM Unitarity Triangle\\
		28  November - 3 December 2016\\
		Tata Institute for Fundamental Research (TIFR), Mumbai, India}

\begin{document}

\section{Introduction}
Mixing between $D^0$ and $\bar{D^0}$ provides crucial information about 
electroweak interactions and the Cabibbo-Kobayashi-Maskawa (CKM) matrix. 
Phenomenon of mixing can be described as decaying two-component quantum states.
\begin{equation}
  \textrm{Mass eigenstates} (D_1, D_2) \neq \textrm{Flavor eigenstates} (D^0, \bar{D}^0). 
\end{equation}
The two parameters characterizing $D^0 - \bar{D}^0$ mixing are
\begin{equation}
  x \equiv \frac{\Delta M}{\Gamma},~~~~~ \Delta M \equiv M_1 - M_2,
\end{equation}

\begin{equation}
  y\equiv \frac{\Delta \Gamma}{2\Gamma},~~~~~ \Delta\Gamma \equiv \Gamma_1 - \Gamma_2,
\end{equation}
where $M_{1,2}$ ($\Gamma_{1,2}$) are the masses (decay widths) of $D_{1,2}$, and $\Gamma \equiv (\Gamma_1+\Gamma_2)/2$ is the mean decay width.

Flavor eigenstates can be written as:
\begin{equation}
  \ket{D^0(t)} = \frac{1}{2p}[\ket{D_1(t)} + \ket{D_2(t)}]~\textrm{and}~\ket{\bar{D}^0(t)} = \frac{1}{2q}[\ket{D_1(t)} - \ket{D_2(t)}]
\end{equation}
The coefficients $p$ and $q$ are complex coefficients satisfying $|p|^2 + |q|^2 =1$, and $q/p = |q/p|e^{i \phi}$.

In the Standard Model (SM), $D^0$-$\bar{D}^0$ mixing is well described by box diagram
containing down-type ($d,s,b$) quarks. While both $s$ and $d$ box amplitudes  
are suppressed by a factor $(m_s^2 - m_d^2)^2/(m_W^2m_c^2)$ due to the
Glashow-Iliopoulos-Maiani  mechanism~\cite{gim},  the contributions from loops 
involving $b$ quarks is further suppressed by CKM factors 
$|V_{ub}V^*_{cb}|^2/|V_{us}V_{cs}^*|^2 =\mathcal{O}(10^{-6})$. The short-distance
SM predictions are $x = \mathcal{O}(10^{-5})$ and $y=\mathcal{O}(10^{-7})$~\cite{bigi,falk,carlos}.
The long-distance contributions can yield $x,y \leq 10^{-3}$~\cite{carlos}. 
Further, $SU(3)_F$ violation in the final-state phase space could provide 
enough breaking to generate $y \sim 10^{-2}$~\cite{falk2} and 
$x \sim 10^{-3}-10^{-2}$~\cite{falk,carlos}. 
New Physics (NP) can enhance the $D^0$-$\bar{D}^0$ mixing rate~\cite{petrov,golowich}.
Currently, $D^0$-$\bar{D}^0$ mixing has been observed and well established.
Due to the uncertainties in both SM and NP, observation 
at $\mathcal{O}(10^{-2})$ does not indicate presence of NP~\cite{carlos}.

$CP$ violation (CPV)  can play an important role in search for NP. In $D$ meson 
decays, it is categorized as:
\begin{itemize}
\item{CPV in mixing occurs when the mixing probability of $D^0$ to $\bar{D}^0$
is different than that of $\bar{D}^0$ to $D^0$. This happens 
if and only if $|q/p|\neq 1$. Depends only on the mixing parameters 
and not on the final state of decay.}
\item{Direct CPV appears when the amplitude for a decay and its
$CP$ conjugate processes have different magnitudes. It occurs only 
  if the differences between $CP$-conserving strong phases and the
  differences between the $CP$-violating weak phases of the two contributing 
  amplitudes are non-zero.}
\item{CPV in the interference between a direct decay $D^0 \to f$, and decay 
  involving  mixing, $D^0 \to \bar{D}^0 \to f$.}
\end{itemize}

The SM predicts $CP$ asymmetries in $D$ meson to be very small, less than 
$\mathcal{O}(0.01\%)$~\cite{buccella,bianco,petrov2,bobrowski}. NP scenarios such as supersymmetric gluino-squark 
loops, yield direct $CP$ asymmetries as large as $\mathcal{O}(1\%)$~\cite{grossman}.

\section {Mixing results from {\boldmath $B$}-factories}
\subsection{Wrong sign decay {\boldmath $D^0 \to  K^+ \pi^-$}}
In wrong sign (WS) $D^0$ decay $D^0 \to K^+ \pi^-$, the final state is reached 
either through direct doubly Cabibbo suppressed (DCS) decay, or via
mixing where $D^0 \to \bar{D}^0$ and then $\bar{D}^0 \to K^+ \pi^-$ through 
Cabibbo favored (CF) right sign (RS) decay. Interference between the two 
amplitude occurs. One can normalize the WS  to the RS rate to obtain
\begin{equation}
R(t) = \frac{N_{\rm WS}(t)}{N_{\rm RS}(t)} = R_D + \sqrt{R_D}~ y'~ \Gamma t + 
\frac{x'^2 + y'^2}{4}~(\Gamma t)^2,
\end{equation}
where $R_D=|\frac{A_{\rm DCS}}{A_{\rm CF}}|^2$, and the 
$x'$ and $y'$ are related to the mixing
parameters ($x$ and $y$) through a rotation by the strong phase, $\delta_{K\pi}$:
\begin{equation}
  x'\equiv x~ \cos \delta_{K\pi} + y~ \sin  \delta_{K\pi},
\end{equation}
\begin{equation}
  y'\equiv y~ \cos \delta_{K\pi} - x~ \sin  \delta_{K\pi}.
\end{equation}

\begin{table*}[!htbp]
\caption{ Mixing parameters measured by different experiments.
  The quoted uncertainties include both statistical and systematic. 
 }
\begin{center}

  \begin{tabular}{lcccc}
\hline \hline
Experiment & $R_D , \times 10^{-3}$ & $y', \times 10^{-3}$ &$x'^2 , \times 10^{-3}$  \\ \hline
Belle \cite{belle1} &$3.53 \pm 0.13$ & $4.6\pm3.4$ & $0.09\pm0.22$ \\ \hline
BaBar  \cite{babar1}&$3.03 \pm 0.19$ & $9.7\pm5.4$ & $-0.22\pm0.37$ \\ \hline
CDF \cite{cdf1} &$3.51 \pm 0.35$ & $4.3\pm4.3$ & $0.08\pm0.18$ \\ \hline
LHCb \cite{lhcb1} &$3.533 \pm 0.054$ & $5.23\pm0.84$ & $3.6\pm4.3$ \\ \hline

\end{tabular}
\label{tab:1}
\end{center}
\end{table*}

The relative WS decay rate at $B$-factories allows a determination of
$x'^2$, $y'$ and $R_D$, but
not the strong phase $\delta_{K\pi}$.

$B$-factories use slow pion $\pi_s^+$ of the strong decay  $D^{*+} \to D^0 \pi^+$
to determine (`tag') the charm flavor. The charge of $\pi_s$ and the charge of kaon from
decay products of $D$ is used to identify the WS and RS. The values of $x'^2$ and $y'$ are
extracted by a fit to the time-dependent ratio of WS to RS 
decay. Belle~\cite{belle1} (BaBar~\cite{babar1}) excluded  non-mixing 
hypothesis at 5.1 $\sigma$ (3.9 $\sigma$). Table~\ref{tab:1} summarizes the 
mixing parameters by different experiments. Belle observed the mixing using
WS $D$ decay.

\subsection{Decays to $CP$ eigenstates {\boldmath $D^0\to K^+K^-/\pi^+\pi^-$}}
Mixing in $D^0$ decays to $CP$ eigenstates, gives rise to an effective lifetime $\tau$ that differs from that in the decays to flavor eigenstates such as 
$D^0\to K^-\pi^+$. The mixing parameter $y$ can thus be measured by comparing the 
rate of $D^0$ decays to $CP$ eigenstates with decays to non-$CP$ eigenstates. If 
decays to $CP$ eigenstates have a shorter effective lifetime than those decaying
to non-$CP$ eigenstates, then $y$ would be positive~\cite{carlos}. Belle~\cite{bellekk} has measure:
\begin{equation}
y_{CP}= [+1.11\pm0.22\pm0.09]\%~{\rm and}~A_\Gamma = [-0.03\pm0.20\pm0.07]\%
\end{equation}
using 976 $fb^{-1}$ data,  while BaBar~\cite{babarkk} used 468 $fb^{-1}$ to measured:
\begin{equation}
y_{CP}= [+0.72\pm0.18\pm0.12]\%~{\rm and}~ A_\Gamma = [-0.18\pm0.52\pm0.12]\%.
\end{equation}
 The first uncertainty is statistical and the second systematic. The $y_{CP}$
results from Belle~\cite{bellekk} (BaBar~\cite{babarkk}) exclude 
the null mixing hypothesis at 4.7 $\sigma$ (3.3 $\sigma$) significance.

\subsection{Time-dependent analysis of three-body decay modes}
Using amplitude analyses of multi-body $D^0$ decay modes, one can measure 
mixing without the ambiguity of an unknown strong phase. Interferences between
intermediate resonances provide sensitivity to both magnitude and sign of the 
mixing parameters.
Belle and BaBar have performed  mixing studies using $D^0$ decay to 
$K_S^0 \pi^+ \pi^-$ and $K_S^0 K^+ K^-$ final states.  
 
The particle-antiparticle mixing phenomenon causes an initially produced (at 
proper time $t=0$) pure $D^0$ or $\bar{D}^0$ meson state to evolve in time to a
linear combination of $D^0$ and $\bar{D}^0$ states. One can describe the decay
amplitude for $D^0$ ($\bar{D}^0$) into the final state, $\mathcal{A}_f$ 
($\bar{\mathcal{A}}_f$), as a function of Dalitz plot (DP) variables.
Time-dependent decay amplitudes for these decays are:
\begin{equation}
\mathcal{M}(m^2_-, m^2_+,t)=\mathcal{A}(m^2_-, m^2_+)\frac{e_1(t)+e_2(t)}{2}+
\frac{q}{p}\bar{\mathcal{A}}(m^2_-, m^2_+)\frac{e_1(t)-e_2(t)}{2}
\end{equation}
\begin{equation}
\bar{\mathcal{M}}(m^2_-, m^2_+,t)=\bar{\mathcal{A}}(m^2_-, m^2_+)\frac{e_1(t)+e_2(t)}{2}+
\frac{p}{q}\mathcal{A}(m^2_-, m^2_+)\frac{e_1(t)-e_2(t)}{2}
\end{equation}
where $\mathcal{A}$ ($\bar{\mathcal{A}}$) decay amplitude for 
$D^0$ ($\bar{D}^0$), $m_{\pm}^2 \equiv m^2(K_S^0 \pi^\pm)$ is parameterized 
with an amplitude $a_r$ and a phase $\phi_r$, 
$\mathcal{A}(m^2_-, m^2_+) = \sum_{r} a_r e^{i \phi_r}\mathcal{A}_r(m^2_-, m^2_+) + a_{nr} e^{i\phi_{nr}}$ and
$\bar{\mathcal{A}}(m^2_-, m^2_+) = \sum_{r} \bar{a}_r e^{i \bar{\phi}_r}{\mathcal{A}}_r(m^2_-, m^2_+) + \bar{a}_{nr} e^{i \bar{\phi}_{nr}}$.  Time dependence is 
contained in
$e_{1,2}(t) = e^{-i (m_{1,2}-i \Gamma_{1,2}/2)t} $.

In order to fit the DP distribution as function of time, one needs to assume
an amplitude model. These models include a coherent sum of quasi-two-body 
intermediate resonances ($r$) plus a nonresonant ($nr$) component. 
$P$- and $D$-wave 
amplitudes are modeled by Breit-Wigner (BW) or Gounaris-Sakurai functional 
forms, 
including Blatt-Weisskopf centrifugal barrier factors. For describing
$\pi \pi$ $S$-wave
dynamics, the $K$-matrix formalism with $P$-vector approximation is used.

Belle~\cite{belleKspipi} obtained $1231731\pm1633$ signal events for
$D^0 \to K_S^0 \pi^+ \pi^-$ with purity of 95.5\% by using 
$921 fb^{-1}$.  Two observables $M_{K_S^0 \pi^+ \pi^-}$ and $Q\equiv M(K_S^0 \pi^+ \pi^- \pi_S) - M(K_S^0 \pi^+ \pi^-) - m(\pi_s)$ are used to identify the signal.  Using $CP$ conserved fit, Belle measured $ x= (0.56 \pm 0.19^{+0.03+0.06}_{-0.09-0.09})\%$ and $y=(0.30\pm0.15^{+0.04+0.03}_{-0.05-0.06})\%$. No mixing hypothesis is 
excluded  with significance of 2.5$\sigma$. 
Also a search for $CPV$ was carried out measuring 
$|q/p|=0.90^{+0.16+0.05+0.06}_{-0.15-0.04-0.05}$ and arg$(q/p)=(-6\pm11\pm3^{+3}_{-4})^\circ$. The $x$ and $y$ values are consistent with $CP$ conserved fit. 
The last uncertainty is  due to the amplitude model.

BaBar~\cite{babarKshh} used $M_D^0$ and $\Delta M$ to identify the signal and 
obtained
$540800\pm800$ ($79900\pm300$) signal events in the
$D^0 \to K_S^0 \pi^+ \pi^-$  ($D^0 \to K_S^0 K^+ K^-$) decay. Mixing
hypothesis is favored with significance of 1.9$\sigma$. Results
for the nominal mixing fit, in which both $D^0$ and $\bar{D}^0$ samples
from $K_S^0 \pi^+ \pi^-$ and $K_S^0 K^+ K^-$ channels are combined, are 
$x = (1.6\pm2.3\pm1.2\pm0.8)\times 10^{-3}$ and
$y=(5.7\pm2.0\pm1.3\pm0.7) \times 10^{-3}$. 

BaBar also perfromed  the first measurement of mixing parameters from a  
time-dependent 
amplitude analysis of the singly Cabibbo-suppressed  (SCS) decay 
$D^0 \to \pi^+ \pi^- \pi^0$~\cite{babarppp0}. Signal is
identified with the $\Delta M$ variable. Using an isobar model of relativistic BW line shape, 
they measured
$x=(1.5\pm1.2\pm0.6)\%$ and $y=(0.2\pm0.9\pm0.5)\%$. Owing to less statistics, no
$CP$ violation was attempted.

\section{ Direct {\boldmath $CP$} asymmetry measurement}
$D^0$ candidates are selected from the decay $D^{*+} \to D^0 \pi_s^+$,
where $\pi^+_s$ reveals the flavor content of neutral $D$ meson.  The 
$D^{*+}$ momentum calculated in the $e^+ e^-$ center-of-mass frame is used to 
suppress $D^{*+}$  from $B$ decays as well as to reduce the combinatorial
background.  $D^{*+}$ mesons mostly originate from $e^+e^- \to c\bar{c}$ 
process via hadronization, where the inclusive yield has a large uncertainty of
12.5\%~\cite{pdg}. To avoid this uncertainty, we measure the branching fraction of 
signal decay with respect to the well measured mode as normalization 
mode
\begin{equation}
\mathcal{B}_{\rm sig}= \mathcal{B}_{\rm norm} \times \frac{N_{\rm sig}}{N_{\rm norm}} 
\times \frac{\epsilon_{\rm norm}}{\epsilon_{\rm sig}},
\end{equation} 
where $N$ is the extracted yield, $\epsilon$ the reconstructed efficiency and 
$\mathcal{B}$ the branching fraction for signal (sig) and normalization (norm) 
modes. For $\mathcal{B}_{\rm norm}$, the world average values~\cite{pdg} 
is used. Assuming the total decay width to be same for particles and 
antiparticles, the time-integrated $A_{CP}$ is:
\begin{equation}
A_{CP} = \frac{\Gamma(D^0 \to f) - \Gamma(\bar{D}^0 \to \bar{f})}{\Gamma(D^0 \to f) + \Gamma(\bar{D}^0 \to \bar{f})},
\end{equation}
where, $\Gamma$ represents the partial decay width and $f$ is specific final state. The extracted raw asymmetry is given by:
\begin{equation}
A_{\rm raw}= \frac{N(D^0 \to f) - N(\bar{D}^0 \to \bar{f})}{N(D^0 \to f) + N(\bar{D}^0 \to \bar{f})} = A_{CP} + A_{FB} + A_{\epsilon}^{\pi_s}.
\end{equation}
Here, $A_{FB}$ is the forward-backward production asymmetry, and $ A_{\epsilon}^{\pi_s}$ is asymmetry due to difference in detection efficiencies for positively
and negatively charged pions. Both can be eliminated through a relative measurement of $A_{CP}$ if the charged final-state particles are identical. The $CP$ 
asymmetry of the signal mode can then be expressed as:
\begin{equation}
A_{CP}({\rm sig}) = A_{\rm raw}({\rm sig}) - A_{\rm raw}({\rm norm}) + A_{CP}({\rm norm}).
\end{equation}
For the $A_{CP}({\rm norm})$, the world average value~\cite{pdg} is used. This way
one can also reduce systematic uncertainties as those
are common to both the signal and normalization mode get
canceled.

\subsection{{\boldmath $D^0 \to V \gamma$} study}
Radiative charm decays are dominated by non-perturbative long range dynamics, 
so measurements of their branching fractions can be a useful test for the QCD 
based theoretical calculations. Further motivation for a study of 
$D^0 \to V \gamma$, where $V$ is a vector meson, arises due to the potential 
sensitivity of  these decays to NP via $A_{CP}$ measurement. Some
studies predict that $A_{CP}$ can rise to several percent in contrast
to $\mathcal{O}(10^{-3})$ SM expectation~\cite{isidori,lyon}.

Belle~\cite{bellevg} performed the first measurement of $CP$ violation in 
$D^0 \to V\gamma$
decays using 943 $fb^{-1}$ of data.  The signal decays are reconstructed in the sub-decay 
channels of the vector meson: $\phi \to K^+ K^-$, $\bar{K}^{*0} \to K^- \pi^+$ 
and $\rho \to \pi^+ \pi^-$. The corresponding
normalization modes are $D^0 \to K^+ K^-$ ($\phi$ mode), $D^0 \to K^- \pi^+$
($\bar{K}^{*0}$ mode) and $D^0 \to \pi^+ \pi^-$ ($\rho^0$ mode). 

Signal is extracted via a simultaneous two-dimensional fit to the invariant
mass $m(D^0)$ and the cosine of the helicity angle ($\cos\theta_H$), which
is the angle between $D^0$ and one of the charged hadrons in the rest frame of the 
$V$ meson.
We measure: \\
$\mathcal{B}(D^0 \to \phi \gamma) = (2.76\pm 0.19\pm0.10)\times 10^{-5}$,  ~~~
$A_{CP}(D^0 \to \phi\gamma) = -0.094\pm0.066\pm0.001$,\\
$\mathcal{B}(D^0 \to \bar{K}^{*0} \gamma) = (4.66\pm 0.21\pm0.21)\times 10^{-4}$, 
~~~$A_{CP}(D^0 \to \bar{K}^{*0}\gamma) = -0.003\pm0.020\pm0.000$, \\
$\mathcal{B}(D^0 \to \rho^0 \gamma) = (1.77\pm 0.30\pm0.07)\times 10^{-5}$, 
~~~$A_{CP}(D^0 \to \rho^0\gamma) = +0.056\pm0.152\pm0.006$, \\
where the first uncertainty is statistical and the second is systematic. Results
are consistent with no $CP$ asymmetry in any of the $D^0 \to V \gamma$ decay 
modes. Further, the $D^0 \to \rho^0 \gamma$ decay is observed for the first time.

\subsection{{\boldmath $D^0 \to K_S^0 K_S^0$} study}
SCS decays such as $D^0 \to K_S^0 K_S^0$ are of special
interest as possible interference with NP amplitude could lead to larger non-zero
CPV. SM based calculations estimate that direct $CP$ violation in this
decay mode can reach upto 1.1\% (at 95\% confidence level)~\cite{nierste}.
Earlier search for
$CP$ asymmetry in $D^0 \to K_S^0 K_S^0$ has been  performed by the CLEO 
Collaboration 
as $(-23\pm19)\%$~\cite{cleoksks}  and LHCb as $(-2.9\pm5.2\pm2.2)\%$~\cite{lhcbksks}.

Belle extract signal via a simultaneous fit of the $\Delta M$ variable 
using  the normalization mode  $D^0 \to K_S^0 \pi^0$. The signal yield for 
$D^0 \to K_S^0 K_S^0$ is $5,399 \pm 87$ and for $D^0 \to K_S^0 \pi^0$ as
$531,807 \pm 796$ events. A simultaneous fit to the $\Delta M$ distribution of $D^{*+}$ and
$D^{*-}$ is used to estimate the asymmetry.The preliminary 
time-integrated 
$CP$-violating asymmetry $A_{CP}$ obtained using  921 $fb^{-1}$ 
in the $D^0 \to K_S^0  K_S^0$ decay is 
$A_{CP} = (-0.02\pm1.53\pm 0.17)\%$~\cite{belleksks}. 
The dominant 
systematic uncertainty comes from the $A_{CP}$ error of the normalization
channel. The result is consistent with SM expectation and is a significantly
improves over the previous measurements.

\section*{Acknowledgments}
I would like to thank KEKB and other  members of the Belle Collaboration 
along with 
their supporting funding agencies. Further, I would also like to thank the
BaBar Collaboration. This work is supported by INSPIRE Faculty Award from the 
Department of Science and Technology, India.


\begin{thebibliography}{99}
\bibitem{gim} S.L. Glashow, J. Iliopoulos, and L. Maiani, \Journal{\PRD}{2}{1285}{1970}.
\bibitem{bigi} I.I. Bigi and N.G. Uraltsev, \Journal{Nucl. Phys. B}{592}{92}{2001}.
\bibitem{falk} A.F. Falk, Y. Grossman, Z. Ligeti, and A.A. Petrov, \Journal{\PRD}{69}{114021}{2004}.
\bibitem{carlos} C.A. Chavez, R.F. Cowan, and W.S. Lockman, \Journal{Int. J. Mod. Phys. A}{27}{1230019}{2012}.
\bibitem{falk2} A.F. Falk {\it et al.}, \Journal{\PRD}{65}{054034}{2002}.
\bibitem{petrov} A.A. Petrov, \Journal{Int. J. Mod. Phys. A}{21}{5686}{2006}.
\bibitem{golowich} E. Golowich, J. Hewett, S. Pakvasa, and A.A. Petrov, \Journal{\PRD}{76}{095009}{2007}.
\bibitem{buccella} F. Buccella {\it et al.}, \Journal{\PRD}{51}{3478}{1995}.
\bibitem{bianco} S. Bianco, F.L. Fabbri, D. Benson, and I. Bigi,  \Journal{Riv. Nuovo Cimento}{26N7}{1}{2003}.
\bibitem{petrov2} A.A. Petrov, \Journal{\PRD}{69}{111901(R)}{2004}.
\bibitem{bobrowski} M. Bobrowski {\it et al.}, \Journal{JHEP}{03}{009}{2010}.
\bibitem{grossman}Y. Grossman, A.L. Kagan and Y. Nir, \Journal{\PRD}{75}{036008}{2007}.


\bibitem{belle1} B.R. Ko {\it et al.}  (Belle Collaboration),  \Journal{\PRL}{112}{111801}{2014}.
\bibitem{babar1} B. Aubert {\it et al.}  (BaBar Collaboration),  \Journal{\PRL}{98}{211802}{2007}.
\bibitem{cdf1} T. Aaltonen {\it et al.}  (CDF Collaboration),  \Journal{\PRL}{111}{231802}{2013}.
\bibitem{lhcb1} R. Aaij {\it et al.}  (LHCb Collaboration), LHCb-PAPER-2016-033, arXiv:1611.06143 [hep-ex]. 
\bibitem{bellekk} M. Staric {\it et al.} (Belle Collaboration), \Journal{\PLB}{753}{412}{2015}.
\bibitem{babarkk} J.P. Lees {\it et al.} (BaBar Collaboration), \Journal{\PRD}{87}{012004}{2013}.
\bibitem{belleKspipi}T. Peng {\it et al.} (Belle Collaboration), \Journal{\PRD}{89}{091103(R)}{2014}.
\bibitem{babarKshh} P. del Amo Sanchez  {\it et al.}  (BaBar Collaboration), \Journal{\PRL}{105}{081803}{2010}.
\bibitem{babarppp0} J.P. Lees {\it et al.} (BaBar Collaboration), \Journal{\PRD}{93}{112014}{2016}.
\bibitem{pdg} C. Patrignani {\it et al.} (Particle Data Group), Chin. Phys. C, {\bf 40}, 100001 (2016).

\bibitem{isidori} G. Isiodri, and J.F. Kamenik, \Journal{\PRL}{109}{171801}{2012}.

\bibitem{lyon} J. Lyon, and R. Zwicky, Edinburgh/12/16. DIAS-2012-24, arXiv:1210.6546 [hep-ph].

\bibitem{bellevg} T. Nanut {\it et al.} (Belle Collaboration), \Journal{\PRL}{118}{051801}{2017}.
\bibitem{nierste} U. Nierste and S. Schacht, \Journal{\PRD}{92}{054036}{2015}.
\bibitem{cleoksks} G. Bonvicini {\it et al.} (CLEO Collaboration), \Journal{\PRD}{63}{071101}{2001}.
\bibitem{lhcbksks} R. Aaij {\it et al.} (LHCb Collaboration), \Journal{JHEP}{10}{055}{2015}.
\bibitem{belleksks}  A. Abdesselam {\it et al.} (Belle Collaboration), BELLE-CONF-1609, arXiv:1609.06393 [hep-ex].


\end{thebibliography}
\end{document}